\documentclass[twocolumn,amsmath,amssymb,floatfix,superscriptaddress]{revtex4}
\usepackage{graphicx}
\usepackage{dcolumn}
\begin{document}
\title{Aging and Energy Landscapes: 
Application to Liquids and Glasses }
\author{Stefano Mossa} 
\altaffiliation[Present address: ]
{Laboratoire de Physique Theorique des Liquides,
Universite Pierre et Marie Curie , 4 place Jussieu, 
Paris 75005, France}
\affiliation{Dipartimento di Fisica and INFM Udr and 
Center for Statistical Mechanics and Complexity,
Universita' di Roma "La Sapienza" 
Piazzale Aldo Moro 2, I-00185, Roma, Italy}
\affiliation{Center for Polymer Studies and Department of Physics,
Boston University, Boston, Massachusetts 02215}
\author{Emilia La Nave}
\affiliation{Dipartimento di Fisica and INFM Udr and 
Center for Statistical Mechanics and Complexity,
Universita' di Roma "La Sapienza" 
Piazzale Aldo Moro 2, I-00185, Roma, Italy}
\author{Francesco Sciortino}
\affiliation{Dipartimento di Fisica and INFM Udr and 
Center for Statistical Mechanics and Complexity,
Universita' di Roma "La Sapienza" 
Piazzale Aldo Moro 2, I-00185, Roma, Italy}
\author{Piero Tartaglia}
\affiliation{Dipartimento di Fisica and INFM Udr and 
Center for Statistical Mechanics and Complexity,
Universita' di Roma "La Sapienza" 
Piazzale Aldo Moro 2, I-00185, Roma, Italy}
\date{\today}
\begin{abstract}
The equation of state for a liquid in equilibrium,
written in the potential energy landscape formalism,
is generalized to describe  out-of-equilibrium
conditions. The hypothesis that during aging the system explores
basins associated to equilibrium configurations
is the key ingredient in the derivation.
Theoretical predictions are successfully compared with data 
from molecular dynamics simulations of different aging processes, 
such as temperature and pressure jumps.
\end{abstract}
\maketitle
Glasses are out-of-equilibrium (OOE) systems, 
characterized by slow dynamical evolution (aging) 
and history dependent properties. Their thermodynamic
and dynamic properties depend on the preparation 
method (cooling or compression schedules) as well 
as the time spent in the glass phase~\cite{pablobook}. 
The extremely slow aging dynamics has often been considered 
an indication that a thermodynamic description 
of the glass state can be achieved by adding one or more history-dependent
parameters to the  equation of state (EOS)
~\cite{tool,davies,kovacs,McKenna,speedy,kurchan,teo,angellmckenna,franz,leuzzi}. 
Recent theoretical work --- mostly based on
mean field models of structural glasses, where analytic solutions of the
OOE dynamics can be explicitly worked out~\cite{cugliandolo} --- supports
such possibility.

In this Letter we derive an EOS for OOE conditions,
based on a generalization of  the potential energy landscape (PEL) 
thermodynamic approach~\cite{sw,debenedetti01,sastry2001}. 
The EOS is derived under the hypothesis that during aging  
the system explores basins associated with equilibrium configurations, 
a condition which, as shown later, is simple to implement in the PEL
context. The proposed EOS for OOE conditions 
depends on one additional parameter, which for example can be chosen to be
the average depth of the explored local minima of the PEL 
(the so-called inherent structures (IS)~\cite{sw,debenedetti01}). 
The EOS allows for the first time a detailed comparison between  
predictions and ``exact results'' from  
out-of-equilibrium molecular dynamics simulations. 
We report such a comparison for a realistic model  
of the fragile liquid orthoterphenyl, one of the most studied
glass forming liquids~\cite{toelle}. 

The PEL formalism~\cite{sw,debenedetti01} allows for a
clear separation of vibrational and configurational  
contributions in thermodynamical quantities. 
Indeed, the equilibrium liquid free energy $F(T,V)$ at temperature $T$ and volume $V$ 
is expressed as a sum of a vibrational contribution $f_{vib}$, 
representing the intra-basin free energy~\cite{sw},
and a configurational contribution ($-TS_{conf}+e_{IS}$).
Here the configurational entropy $S_{conf}$
accounts for the number of explored basins of the PEL 
of average energy depth $e_{IS}$. $f_{vib}$, for a system of $N$ atoms,  
depends on the curvature of the potential 
energy surface, i.e., on  the $3N$ eigenvalues $\omega_i^2$ of  the
Hessian matrix evaluated at the IS.  Recently,
calculating the $V$ derivative of $F(T,V)$, an analytic EOS has
been derived, based only on the statistical properties of the 
landscape~\cite{lanave02,deben}. 

The separation in vibrational and configurational parts can be
carried out for several properties. 
In our case, it is particularly useful to separate the instantaneous 
pressure as sum of two contributions: the pressure felt by the system
in the inherent structure configuration ($P_{IS}$), and an additional 
contribution related to the finite temperature, which 
we refer to as the vibrational 
contribution ($P_{vib}$)~\cite{clarification}
\begin{equation}
P(T,V,e_{IS})=P_{IS}(e_{IS},V)+P_{vib}(T,V,e_{IS}).
\label{eq:pressure}
\end{equation}
In equilibrium $e_{IS}$ is a well defined function of $T$ and $V$ 
and, as well known, $P$ depends only on $T$ and $V$.
Note that $P_{IS}$, being the pressure experienced by the liquid in the
IS, depends only on the basin's depth and on $V$. Instead, $P_{vib}$
depends on the basin's depth as well as on $T$ and $V$~\cite{precisazione}.

For models for which  the statistical properties of the PEL, i.e., 
the distribution of shapes and numbers of 
$IS$~\cite{lanave02,deben} are known,  $P_{IS}$ and $P_{vib}$
can be calculated theoretically. In other cases, $P_{IS}$ and $P_{vib}$
can be calculated numerically. In all cases, 
an landscape-based EOS based on Eq.~(\ref{eq:pressure}) can be derived.

The separation of the pressure in two contributions 
is crucial for the possibility of extending the equilibrium EOS to OOE 
conditions, since it provides a direct way to rebuild the pressure
once $T$, $V$ and $e_{IS}$ are known. In the PEL formalism, 
the hypothesis that the system explores basins associated with 
typical equilibrium configurations, i.e., configurations sampled in
equilibrium conditions, is equivalent to
assuming that the relations --- calculated in equilibrium --- 
linking $P_{IS}$ and $P_{vib}$ to $e_{IS}$ and $V$  
are valid also in OOE conditions.
If this is the case, the knowledge of $e_{IS}$, $T$ and $V$ is
sufficient to calculate both $P_{IS}$, $P_{vib}$ and
their sum $P$ according to Eq.~(\ref{eq:pressure}).
\begin{figure}[t]
\includegraphics[width=.43\textwidth]{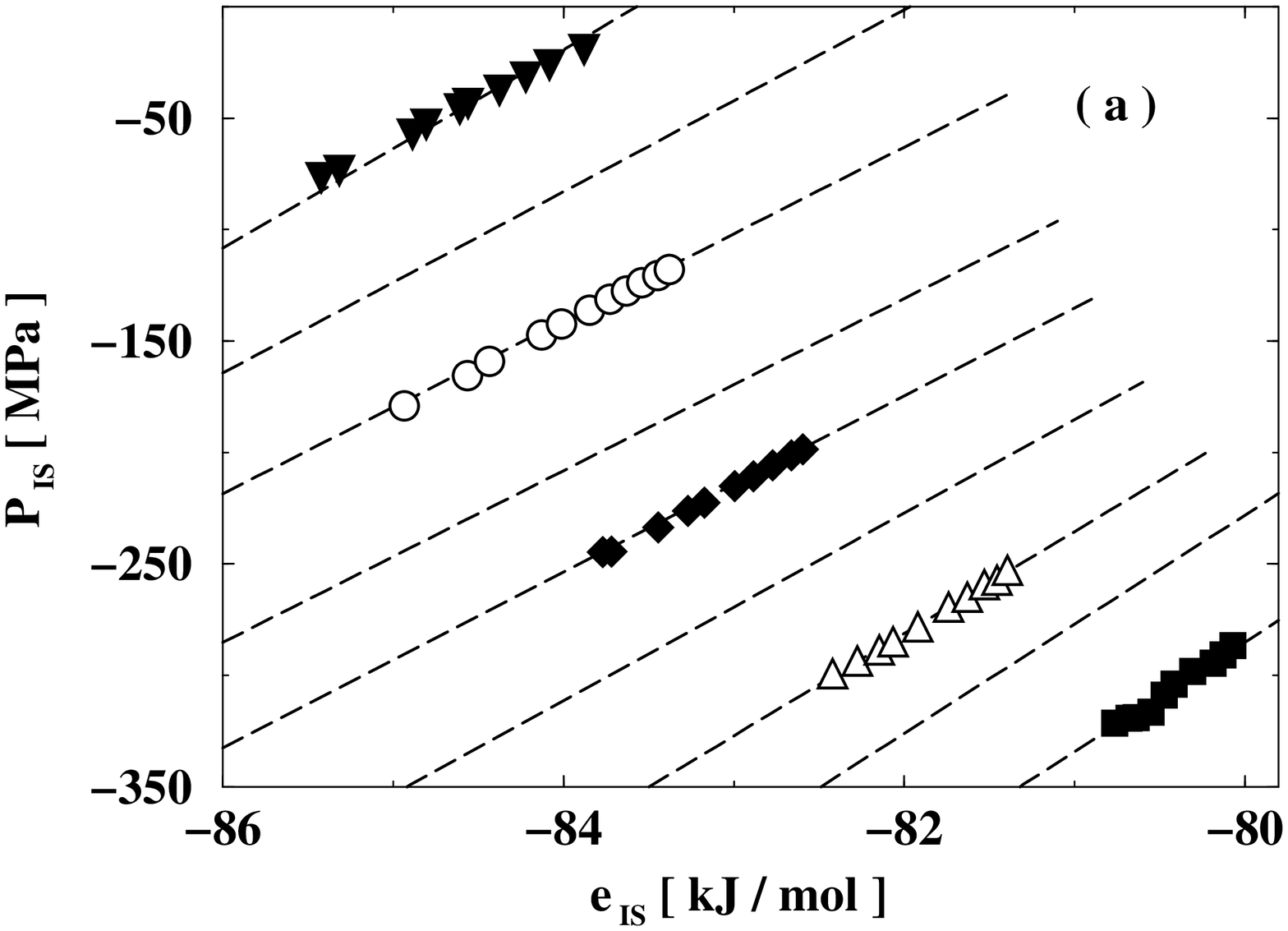}

\includegraphics[width=.43\textwidth]{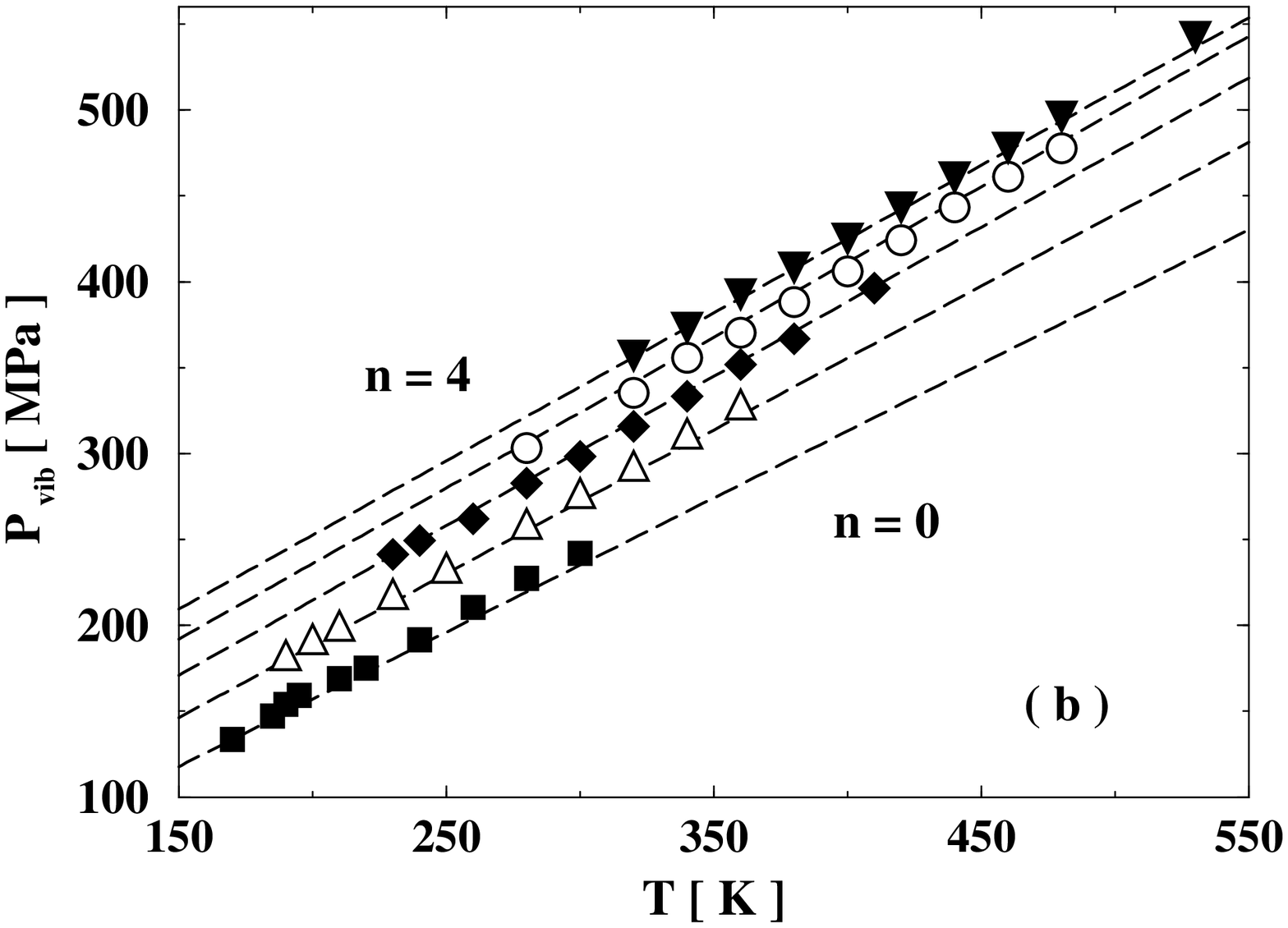}
\vspace{-0.4cm}
\caption{(a) Equilibrium relation between basin pressure 
$P_{IS}$ and basin depth $e_{IS}$. Symbols are 
simulation results at the five volumes per molecule
$0.337$ nm$^3$ (down triangles),
$0.345$ nm$^3$ (circles),
$0.353$ nm$^3$ (diamonds),
$0.361$ nm$^3$ (up triangles), and
$0.369$ nm$^3$ (squares)
from Ref.\protect~\cite{mossa02}. 
Dashed lines are calculated using 
the parameterization  discussed in Ref.\protect~\cite{lanave02}.
Results at four intermediate volumes are also shown.
The local minima are calculated using standard minimization 
algorithms\protect~\cite{mossa02}.
(b) $P_{vib}$ as a function of $T$ for the same five volumes.
Each curve has been shifted by $n \times 20$ MPa
to avoid overlaps. Dashed lines show a suitable parameterization 
of the $e_{IS}$, $V$ and $T$ dependence of $P_{vib}$, 
as discussed in Ref.\protect~\cite{lanave02}.
}
\label{fig:piseis}
\end{figure}
Similarly, the values of $P$, $T$ and $e_{IS}$ are sufficient 
to predict $V$, since both $P_{IS}(e_{IS},V)$ and $P_{vib}(e_{IS},T,V)$ can
be estimated as a function of $V$. The predicted $V$ is the value for
which $P_{IS}(e_{IS},V)+ P_{vib}(e_{IS},T,V)$ matches the external
pressure.

Next we apply the ideas discussed above to a system of
$N=343$ molecules interacting via the Lewis and Wanstr\"{o}m (LW) 
model for the fragile molecular liquid orthoterphenyl 
(OTP)~\cite{mossa02,lewis}.
The LW model is a rigid three-site model, with intermolecular
site-site interactions modeled by the Lennard-Jones (LJ) 
potential~\cite{lewis}. We refer to Ref.~\cite{mossa02} for all 
numerical details. An equilibrium EOS, based on the assumption of a Gaussian
landscape, has been recently presented for this model~\cite{lanave02},
and successfully compared with the ``exact'' EOS calculated 
from molecular dynamics simulations.
\begin{figure}[t]
\centering
\includegraphics[width=.45\textwidth]{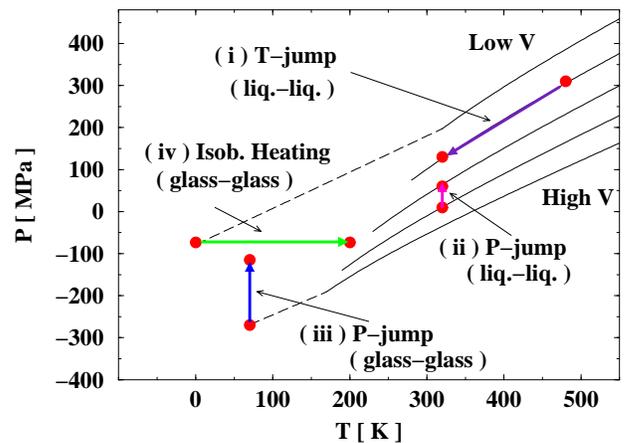}
\vspace{-0.4cm}
\caption{Paths in the $P-T$ plane of the
four OOE simulations discussed in the text. Arrows indicate:
(i) a $T$-jump at constant $V$ between liquid states; 
(ii) a $P$-jump at constant $T$ between liquid states; 
(iii) a $P$-jump at constant $T$ starting from a glass;
(iv) a constant $P$ heating of a glass.
$P$ and $T$-jumps have been simulated by changing the
thermostat and barostat value at $t=0$. The  time constant of the
thermostat and barostat has been fixed to 20~ps.
The full lines indicate the equilibrium $P(T)$ at five $V$ values, from 
Ref.\protect~\cite{mossa02}.
The dashed lines connect the equilibrium state points and the
glass states obtained by fast constant-$V$ cooling.}
\label{fig:paths}
\end{figure}

Fig.~\ref{fig:piseis} shows the relation between $P_{IS}$ and $e_{IS}$
in equilibrium along different isochores and $P_{vib}$ as a function of $T$
for the LW model. Both quantities can be parameterized
in a quite accurate way. Indeed, for this model, $P_{vib}$
is weakly dependent on $V$ and $e_{IS}$ and it is well
represented by a linear $T$-dependence.
The parameterization of $P_{IS}$ and $P_{vib}$ 
offers the possibility of  estimating the system pressure
once $T$, $V$ and $e_{IS}$ are known. 

We study four different OOE protocols via computer simulation.
The imposed external conditions are illustrated in Fig.~\ref{fig:paths}.
We study
{\em i)} a $T$-jump  at constant $V$; 
{\em ii)} a $P$-jump at constant $T$; 
{\em iii)} a $P$-jump at constant $T$ in the glass phase and
{\em iv)} an isobaric heating of a glass.
For the cases {\em i)} and {\em ii)}, conditions are chosen in such a way 
that within the simulation time window (50 $ns$) the system reaches the
equilibrium state. In all studied cases, averages over 50 
independent simulations have been performed.

\begin{figure}[t]
\centering
\includegraphics[width=.43\textwidth]{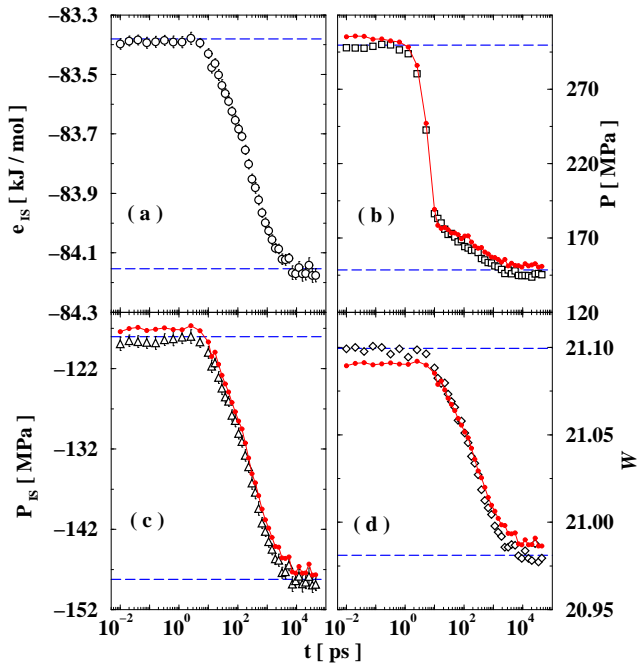}
\vspace{-0.4cm}
\caption{Dynamics after a $T$-jump, at volume per molecule $V=0.345$ nm$^3$,
from $480$ K to $340$ K. (a) $e_{IS}$; (b) $P$; (c) $P_{IS}$; 
(d) $\cal{W}$ ($\omega_o= 1 cm^{-1}$).
Open symbols are simulation results,
closed circles are theoretical predictions based on the OOE PEL-EOS
and using as input the data shown in panel (a).
Dashed lines indicate the known equilibrium values at the initial and
final state.
The error in the prediction at short and long times, when
equilibrium conditions are met, provides a measure of the
quality of the equilibrium EOS~\cite{lanave02}, which agrees
with the simulation data within $\pm 5 MPa$.}
\label{fig:quench}
\end{figure}

Case {\it (i)}. ---
Fig.~\ref{fig:quench} shows the comparison between
the theoretical predictions and the numerical results 
for the constant-$V$ $T$-jump case. The $e_{IS}$ values, 
calculated from the simulation (Fig.~\ref{fig:quench}(a)), 
are used together with the values of $T$ and $V$ as
input to predict the evolution of $P$ (Fig.~\ref{fig:quench}(b)). 
Figs.~\ref{fig:quench}(c) and (d) also show the comparison 
for $P_{IS}$ and $\cal{W}$. 
${\cal W}  (e_{IS})$, defined as $\sum_{i=1}^{3N} \ln( \omega_i(e_{IS}))$,
provides an indication of the similarity in basin shape
(at the level of the harmonic approximation) between basins explored in
equilibrium and in OOE conditions. 
The quality of the predictions indeed supports the validity of the EOS and
the hypothesis that the basins sampled during aging have the same 
relationship ${\cal{W}}(e_{IS})$ as in equilibrium. 

\begin{figure}[t]
\centering
\includegraphics[width=.43\textwidth]{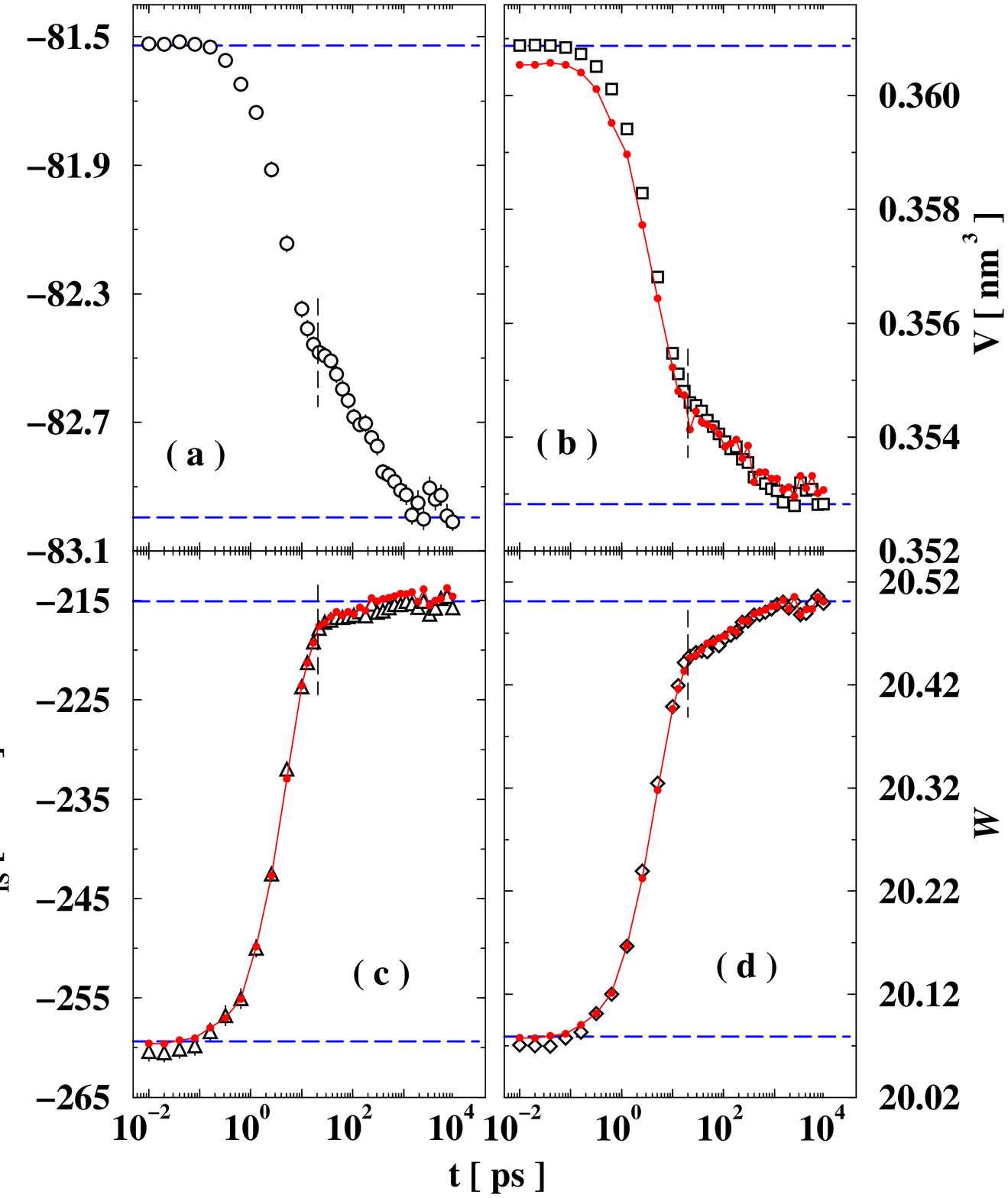}

\includegraphics[width=.43\textwidth]{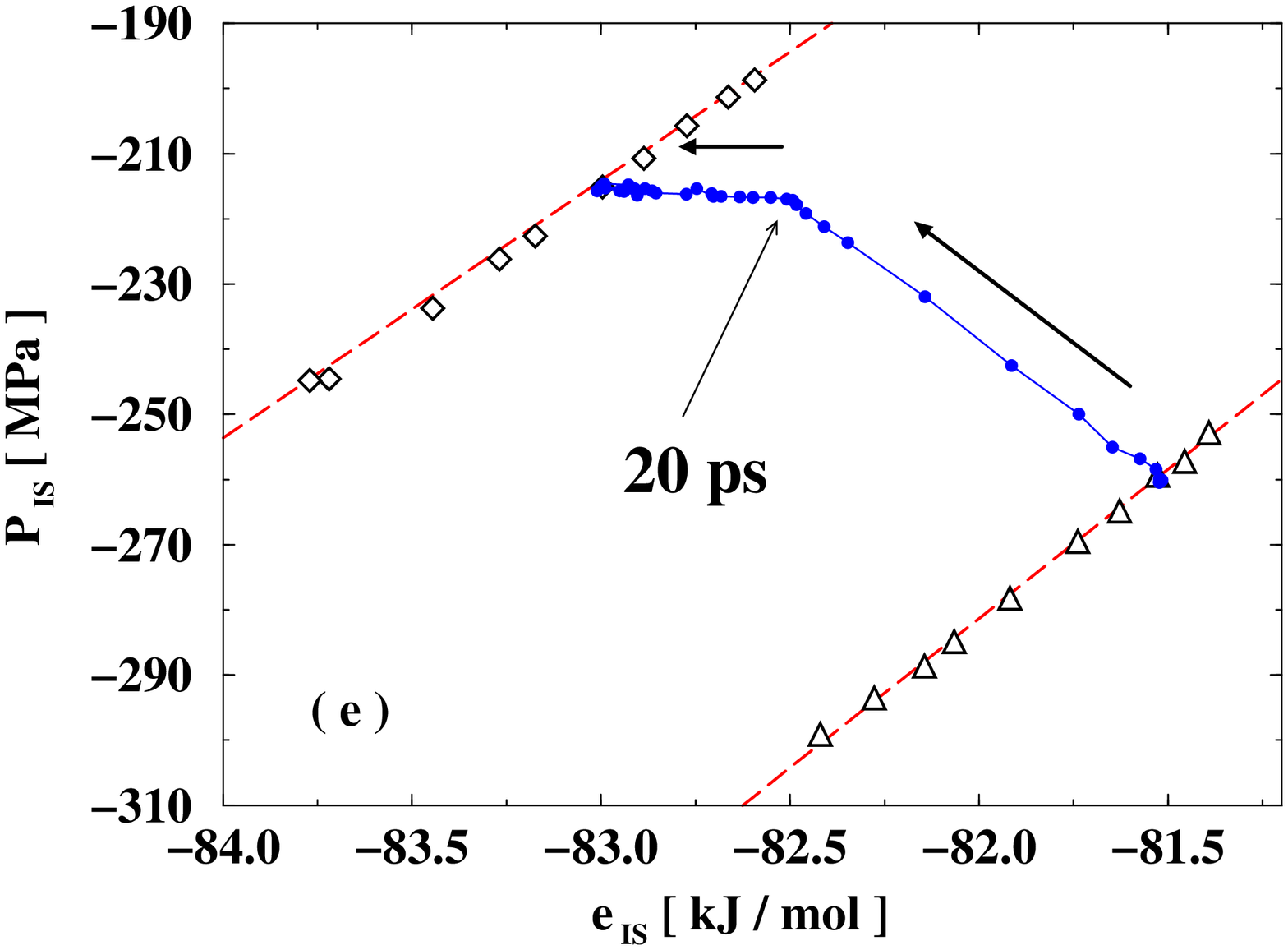}
\vspace{-0.4cm}
\caption{Dynamics after a $P$-jump at $T=320$ K from $13.4$ MPa  to $60.7$ MPa . 
(a) $e_{IS}$; (b) $V$; (c) $P_{IS}$; 
(d) $\cal{W}$.
Open symbols are simulation results,
closed circles are theoretical predictions based on the OOE PEL-EOS, 
using as input the data shown in panel (a).
Horizontal dashed lines indicate the known equilibrium values 
at the initial and final state. The vertical dashed lines 
indicate the time at which the external pressure reaches 
the final value. Panel (e) shows the actual path of the 
aging process in the $P_{IS}$-$e_{IS}$ plane.}
\label{fig:compress}
\end{figure}

Case {\it (ii)}. ---
Fig.~\ref{fig:compress} shows the evolution of the system 
after a constant-$T$ pressure jump.  
In this case, the $e_{IS}$ values calculated from the
simulation are used together with the values of $T$ and $P$
as input to successfully predict the evolution of $V$. 
For times shorter than $20~ps$, the barostat time constant,
the fast increasing external pressure forces the system to change the volume
with a solid-like response. In this time window, the initial basins in
configuration space are continuously deformed by the rapid volume change, 
in agreement with the findings of Ref.~\cite{lacks}.
For time longer than $20~ps$, when the external pressure has reached the
equilibrium value, the system starts to
explore basins different from the (deformed) original ones, and
the time evolution becomes controlled by the aging kinetics.
To support this interpretation we show in Fig.~\ref{fig:compress}(e)
the actual path in the  $e_{IS}$-$P_{IS}$ plane, where the
change around $20~ps$ is clearly evidenced.

\begin{figure}[t]
\centering
\includegraphics[width=.43\textwidth]{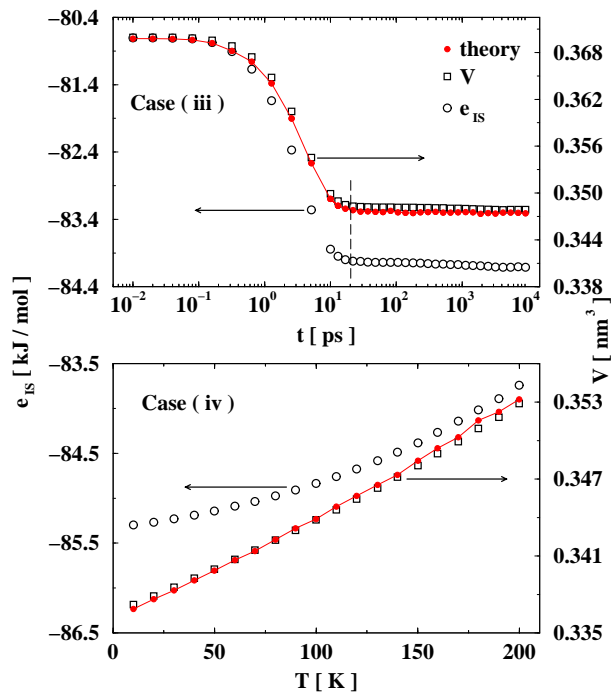}
\vspace{-0.4cm}
\caption{Top: Simulation of a $P$-jump from $-270.1$ MPa 
to $-115.4$ MPa, starting from a glass configuration, 
obtained by quickly cooling to $70$ K 
equilibrium configurations at volume per molecule
$V=0.369$ nm$^3$ and $T=170$ K. 
Bottom: Isobaric heating, starting from the IS at $V=0.337$ 
nm$^3$ per molecule and $T=320$ K. The heating rate is $1$ K / ps. 
Circles are the $e_{IS}$ values used as
input to predict the evolution of $V$ (lines). Squares are the $V$ 
calculated directly from the simulations.}
\label{fig:glass}
\end{figure}

Cases {\it (iii)} and {\it (iv)}. ---
The last two cases are respectively a constant $T$ $P$-jump 
and a heating at constant $P$ (with an heating rate of $1~K/ps$), 
both starting from glass configurations.
These initial configurations are generated by rapid constant-$V$
quenches of equilibrium configurations, as shown schematically in 
Fig.~\ref{fig:paths}.
Fig.~\ref{fig:glass} shows the comparison between the theoretical predictions 
and the numerical calculations for $V$, using $e_{IS}$ as input. 
In the case of a $P$-jump (Fig.~\ref{fig:glass}-Top), again two different dynamical 
behaviors are observed: a fast dynamics process describing 
the mechanical response of the glass to the external pressure change, 
followed by an extremely slow aging dynamics,
during which basin changes are taking place. 
In both cases and for all times, the out-of-equilibrium EOS is able to
predict quantitatively the $V$ changes.

In summary, we have derived an extension of the PEL equation of state to
model the thermodynamic properties of liquids under out-of-equilibrium
conditions. Such an extension --- based on the hypothesis that an aging
system evolves through regions of configuration space which are typically
sampled in equilibrium --- requires an additional thermodynamic parameter
which, for convenience, we have associated with the IS energy of the
explored basins.
 
The results reported in this Letter 
(Figs.~\ref{fig:quench}-\ref{fig:glass}) 
show that the proposed generalization of the PEL 
equation of state is successful in predicting out-of-equilibrium
thermodynamics, at least under the conditions and time scales probed by
state-of-the-art numerical simulations.
Larger $T$ or $P$ jumps and/or longer aging time could require a more
detailed thermodynamic description, with more than one
additional parameter. Despite such a possibility,
the energy landscape approach developed in this Letter is
a promising starting point for looking into more complex problems in 
the physics of OOE liquids and glasses.

We thank A.~Angell, P.~Debenedetti, T.~Keyes,
L.~Leuzzi, G.~McKenna, G.~Ruocco, S.~Sastry, A.~Scala,
and M.~Yamada for discussions and comments.
We acknowledge support from MIUR-COFIN 2000, INFM-PRA and INFM 
Initiative Parallel Computing. 
%
%
%%%%%%%%%%%%
% REFERENCES
%%%%%%%%%%%%
%

%
%

\begin{thebibliography}{99}
%
\bibitem{pablobook} P.~G. Debenedetti,  
{\it Metastable liquids} (Princeton Univ. Press, Princeton, 1996).
% 
\bibitem{tool} A.~Q. Tool, 
J. Am. Ceram. Soc. {\bf 29}, 240 (1946).
%
\bibitem{davies} R.~O. Davies and G.~O. Jones,
Adv. in Physics {\bf 2}, 370 (1953). 
%
\bibitem{kovacs} A.~J. Kovacs, 
Fortschr. Hochpolym.-Forsch. {\bf 3}, 394 (1964). 
%
\bibitem{McKenna} G.~B. McKenna, 
in {\it Comprehensive Polymer Science}, Vol. 2, 
ed. by C. Booth and C. Price, Pergamon, Oxford, 
311 (1989).
%
\bibitem{speedy} R.~J. Speedy,
J. Chem. Phys. {\bf 100}, 6684 (1994).
%
\bibitem{kurchan} L.~F. Cugliandolo, {\it et al.},
%, J. Kurchan, and L. Peliti, 
Phys. Rev. E {\bf 55}, 3898 (1997).
%
\bibitem{teo} Th.~M. Nieuwenhuizen, 
Phys. Rev. Lett. {\bf 80}, 5580 (1998).
%
\bibitem{angellmckenna} 
A.~C. Angell, {\it et al.},
%, K.~L. Ngai, G.~B. McKenna, P.~F McMillan, and S.~W. Martin, 
J. Appl. Phys. {\bf 88}, 3113 (2000).
%
\bibitem{franz}
S. Franz and M.~A. Virasoro,  
J. Phys. A {\bf 33}, 891 (2000).
%
\bibitem{leuzzi} L. Leuzzi and Th.~M. Nieuwenhuizen,
J. Phys.: Condens. Matter {\bf 14}, 1637 (2002). 
%
\bibitem{cugliandolo} L.~F. Cugliandolo, {\it et al.},
%, J. Kurchan, and P. Le Doussal, 
Phys. Rev. Lett. {\bf 76}, 2390 (1996).
%
\bibitem{sw} F.~H. Stillinger and T.~A. Weber, 
Phys. Rev. A {\bf 25}, 978 (1982). 
%
\bibitem{debenedetti01}
P.~G. Debenedetti and F.~H. Stillinger, 
Nature {\bf 410}, 259 (2001).
%
\bibitem{sastry2001}
S. Sastry, Nature {\bf 409}, 164 (2001).
%
\bibitem{toelle} A. T\"olle, 
Rep. Prog. Phys. {\bf 64}, 1473 (2001).
%
\bibitem{lanave02} E. La Nave, {\it et al.},
Phys. Rev. Lett. {\bf 88}, 225701 (2002).
%
\bibitem{deben} P.~G. Debenedetti, {\it et al.}, 
%T.~M. Truskett, C.~P. Lewis and F. Stillinger, 
Adv. in Chem. Eng. {\bf 28}, 21 (2001).
%
\bibitem{clarification}
In equilibrium, $P$ can be unambiguously separated in 
a configurational ${\partial}_V  [T S_{conf}(T,V) - e_{IS}(T,V)]$ 
and a vibrational  ${\partial}_{V } f_{vib}$  
contribution. In the IS configuration, the vibrational part is
strictly zero. Still, the configurational component may not
coincide with $P_{IS}$. Different proposals for $P_{IS}$ 
arise depending on the meaning of $T$ in the $-T S_{conf}$ term above.
See Refs.~\protect\cite{lanave02,st2001}. Such ambiguity 
affects the theoretical expression for $P_{IS}$ 
based on the statistical properties of the landscape. Luckily, 
this ambiguity does not hamper the possibility of developing an
OOE-EOS when $P_{IS}$ is calculated directly from the
simulation data.
%
\bibitem{precisazione}
For example, in an harmonic solid, $P_{vib}$ 
is related to the $V$ derivative of the eigenfrequencies as 
$P_{vib}(T,V,e_{IS})\equiv  k_B T {\partial}_{V } 
\sum_{i=1}^{3N-3} ln(\beta \hbar \omega_i(e_{IS}))$. 
%
\bibitem{mossa02}
S. Mossa, {\it et al.},
Phys. Rev. E {\bf 65}, 041205 (2002).
%
\bibitem{lewis}
L.~J. Lewis and G. Wahnstr\"{o}m, 
Phys. Rev. E {\bf 50}, 3865 (1994).
%
\bibitem{lacks} 
D.~L. Malandro and D.~J. Lacks, 
Phys. Rev. Lett. {\bf 81}, 5576 (1998).
%
\bibitem{st2001} F.~Sciortino and P.~Tartaglia,
Phys. Rev. Lett. {\bf 86}, 107 (2001).
%
\end{thebibliography}
\end{document}